\documentclass[preprint,showpacs,aps,preprintnumbers,amsmath,amssymb,groupedaddress]{revtex4}

\usepackage[dvips]{graphicx}
\usepackage{dcolumn}
\usepackage{bm}
\usepackage{hyperref}

\begin{document}
\preprint{APS/123-QED}

\title{Astrophysics from data analysis of spherical gravitational wave detectors}
\email{chlenzi@ita.br, 050003@cefetsp.br}

\author{C\'{e}sar H. Lenzi$^{1*}$, Nadja S. Magalh\~{a}es$^2$, C\'{e}sar A. Costa$^{3\ddagger}$, Rubens M. Marinho$^{1\dagger}$, Helmo A. B. Ara\'{u}jo$^1$ and Odylio D. Aguiar$^{3}$}%
\affiliation{$^1$Departamento de F\'{i}sica, Instituto
Tecnol\'{o}gico de  Aeron\'{a}utica, Centro T\'{e}cnico
Aeroespacial, Pra\c{c}a Marechal Eduardo Gomes 50, S\~{a}o Jos\'{e}
dos Campos,
SP, 12228-900, Brazil \\
$^2$Centro Federal de Educa\c{c}\~{a}o Tecnol\'{o}gica de S\~{a}o Paulo, R. Dr. Pedro Vicente 625, S\~{a}o Paulo, SP, 01109-010, Brazil \\
$^3$Departamento de Astrof\'{i}sica, Instituto Nacional de Pesquisas
Espaciais,  Av. dos Astronautas 1.758,  S\~{a}o Jos\'{e} dos Campos,
SP, 12227-010, Brazil}
\date{\today}

\begin{abstract}

The direct detection of gravitational waves will provide valuable
astrophysical information about many celestial objects. Also, it
will be an important test to general relativity and other theories
of gravitation. The gravitational wave detector SCHENBERG has
recently undergone its first test run. It is expected to have its
first scientific run soon. In this work the data analysis system of
this spherical, resonant mass detector is tested through the
simulation of the detection of gravitational waves generated during
the inspiralling phase of a binary system. It is shown from the
simulated data that it is not necessary to have all six transducers
operational in order to determine the source's direction and the
wave's amplitudes.
\end{abstract}

\pacs{95.55.Ym, 04.80.Nn, 04.30.-w}

\keywords{Gravitational Waves \and Spherical gravitational wave
detector \and Inverse Problem Solution \and Method of independent
bars}

\maketitle

\section{Introduction}
\label{intro} As predicted by the theory of relativity and other
theories of gravitation, time-dependent gravitational forces are
expected to propagate in spacetime in the form of waves
\cite{mis73}. Strong gravitational waves (g.w.) are expected to be
generated by astrophysical objects \cite{thornebook}. For instance,
two very massive stars orbiting each other would emit such waves. In
particular, when they are coalescing they emit waves in a large
frequency range. Another example is given by a black hole that rings
down, also emitting in different wavelengths \cite{ces04}.

There is the belief within the international community that works
with g.w. detection that the first direct detection of gravitation
radiation from an astrophysical source will become a reality in the
near future. This will open a new window to the universe, bringing
new information about known objects and about fairly unknown things,
like dark matter. In order to extract such information from the huge
amount of data that is expected to be generated from the detectors'
outputs (which has already started) a lot of work will be demanded
in the field of data analysis.

From the analysis of the detected waves emitted by such sources
important information is expected. The very first direct detection
will provide a test for one of the predictions of the theory of
general relativity. Then, continuous observation will allow testing
other theories of gravitation \cite{maga96,merk98,stel06}, besides
initiating gravitational astronomy \cite{maga95,schu05}.

In order to make astrophysical observations the following parameters
that characterize the g.w. are needed: the amplitudes of the two
states of polarization of the wave as functions of time ($h_+ (t)$
and $h_{\times}(t)$) and the source's direction in the sky (given by
the angles $\Theta$ and $\Phi$) \cite{Gursel}. Therefore,
gravitational wave observatories must be able to detect four
independent observable in order to allow for gravitational wave
astronomy to start.

The basis for the existence of resonant-mass gravitational wave
detectors like SCHENBERG\cite{amaldi6} is the fact that solid bodies
are distorted in the presence of g.w. due to the changes in
spacetime. In principle, the spherical geometry of SCHENBERG's
antenna implies no preferred direction of observation (i.e., it is
omnidirectional) and the all observable needed for gravitational
astronomy can be obtained from only one spherical detector
appropriately equipped \cite{maga95,maga97b}. When fully
operational, this detector will be able not only to acknowledge the
presence of a g.w. within its bandwidth: it will be able to inform
the direction of a source in sky, the wave's amplitude and the
polarization components in the detector bandwidth - one only antenna
working as a gravitational wave observatory in a bandwidth between
$3000$Hz and $3400$Hz, sensitive to displacements around
$10^{-20}$m. To this end six transducers will continuously monitor
radial displacements of the antenna's surface. SCHENBERG is
installed at the Physics Institute of the University of S\~{a}o
Paulo (S\~{a}o Paulo city, Brazil) and has undergone its first test
run in September 8, 2006, with three transducers tuned to the above
band.

There are mathematical models for this detector for the case that
all these six transducers are operational. Such models have been
investigated two situations: one in which the transducers are
perfectly uncoupled \cite{maga97b,merk97,cesar06} and another in
which the transducers are somehow coupled to each other
\cite{merk99}. For both cases it is possible to solve the inverse
problem and obtain the four astrophysical parameters needed if one
assume General Relativity as the correct theory of gravitation.

This work presents the results of an investigation
\cite{Cesar_thesis} carried on by the data analysis group within the
GRAVITON project \cite{gravitonhome} (the one SCHENBERG is part of)
aimed at developing a model of the detector with {\it less} than six
transducers. This is an important issue, since occasionally not all
six resonators may be operational simultaneously. In this case there
is a break in the convenient symmetry among the transducers. Two
approaches were under investigation: one considered the model
already developed for 6 transducers \cite{cesartese} and simply
reduced their number; the other, described here, considered the
fewer transducers as independent devices and redesigned the
mathematical model. It was found that the system sphere plus
transducers is equivalent to a system of independent bar detectors
with center-of-masses located at the sphere's center-of-mass
(``cyclical" bars) \cite{maga95}. It was possible to retrieve the
relevant astrophysical information about a coalescing binary neutron
star system both from the sphere coupled to six transducers and from
a system considering the sphere's outputs as the outputs of
transducers coupled to cyclical bar detectors (which will be called
the ``method of independent bars"). These results are detailed as
follows.

\section{Method of Independent bars}
It has already been shown theoretically by Magalh\~aes et
al.\cite{maga95} that any array of $n$ cyclical bars will respond to
a gravitational wave in a way similar as a non-noisy, high Q
spherical antenna monitored by $n$ transducers. In both cases, the
projection $R^L$ of a g.w. onto a detector can be expressed in terms
of the internal product between two symmetric, traceless
tensors\cite{maga95}:
\begin{equation}
R^L  = \sum\limits_{i,j = 1}^3 {W_{ij} D_{ij}^L }.
\end{equation}
In this expression $R^L$ are the outputs of the transducers
(L=1,2,...,n, where $n$ is the number of transducers positioned on
the sphere's surface or, equivalently, the number of cyclical bar
detectors under consideration). The tensor $W$ contains only
information concerning the gravitational wave, while $D^L$ carries
only information concerning the detector. For a {\it bar} detector,
if $\theta _L$ is the angle the detection point (where the L-th
transducer is located) makes with the local zenith and $\phi _L$ is
the azimuthal angle then one can show that
\begin{small}\begin{equation}\label{eq:Dbar}
D^L = \left[\begin{array}{ccc} \sin^2\theta_L\cos^2\phi_L-\frac 13
&
                \frac 12\sin^2\theta_L\sin2\phi_L               &
                \frac 12\sin2\theta_L\cos\phi_L                    \cr
                \frac 12\sin^2\theta_L\sin2\phi_L               &
                \sin^2\theta_L\sin^2\phi_L-\frac 13              &
                \frac 12\sin2\theta_L\sin\phi_L                                  \cr
                \frac 12\sin2\theta_L\cos\phi_L             &
                \frac 12\sin2\theta_L\sin\phi_L                         &
                \cos^2\theta_L-\frac 13
\end{array}\right].
\end{equation}\end{small}
It is the knowledge of the wave's tensor, $W$, that allows one to
determine all astrophysical parameters, ($h_+ (t)$, $h_{\times}(t)$)
and ($\Theta$, $\Phi$), as was shown in \cite{maga95}. In order to
determine $W_{ij}$ one notices that, from general relativity, this
tensor has six independent terms. It is then necessary to have at
least six equations to determine these six unknowns. One equation is
the traceless condition
\begin{equation}
W_{11} + W_{22} + W_{33} = 0,
 \label{traceless}
\end{equation}
 the other five can be given by
\begin{equation}\label{RL}
    \begin{array}{l}
        D_{11}^Lh_{11} + 2D_{12}^Lh_{12} + 2D_{13}^Lh_{13} +\\
          \ \  \\
         \ \ \ \ \ \ D_{22}^Lh_{22} +2D_{23}^Lh_{23} + 2D_{33}^Lh_{33} = 2R^L. \\
    \end{array}
\end{equation}
with $L = 1,...,5$. As usual, at least five cyclical bars with one
transducer each are needed. The six equations given by
(\ref{traceless}) and (\ref{RL}) can be written in matrix form as
\begin{small}
\begin{equation}\label{solproinv}
\left[\begin{array}{cccccc} D_{11}^1 & 2D_{12}^1 & 2D_{13}^1 &
D_{22}^1 & 2D_{23}^1 & D_{33}^1 \cr D_{11}^2 & 2D_{12}^2 & 2D_{13}^2
& D_{22}^2 & 2D_{23}^2 & D_{33}^2 \cr D_{11}^3 & 2D_{12}^3 &
2D_{12}^3 & D_{22}^3 & 2D_{23}^3 & D_{33}^3 \cr D_{11}^4 & 2D_{12}^4
& 2D_{13}^4 & D_{22}^4 & 2D_{23}^4 & D_{33}^4 \cr D_{11}^5 &
2D_{12}^5 & 2D_{13}^5 & D_{22}^5 & 2D_{23}^5 & D_{33}^5 \cr
   1     &     0     &     0     &     1    &     0     &     1    \cr
\end{array}\right]
\left[
\begin{array}{c}
W_{11} \cr W_{12} \cr W_{13} \cr W_{22} \cr W_{23} \cr W_{33} \cr
\end{array}
\right] = \left[
\begin{array}{c}
R^1 \cr R^2 \cr R^3 \cr R^4 \cr R^5 \cr 0
\end{array}
\right]
\end{equation}
\end{small}
or $DW = R$. As long as the matrix $D$ is square and non-singular,
from the experimental knowledge of $R^L$ and $D^L$ one can obtain
the wave's tensor $W$ through the relation

\begin{equation}
W =D^{-1}R.
 \label{eq:W}
\end{equation}

In order to make sure that the use of a tensor for bars,
(\ref{eq:Dbar}), would yield the solution for the inverse problem
using $R^L$ derived from the data of a spherical detector, a test
was performed. The transducers were positioned as planned in
SCHENBERG, located according to the pentagonal faces of a truncated
icosahedron:
\begin{equation}
\begin{array}{cc}
[(\theta,\phi)] = &
[(79.1877^\circ,120^\circ),(79.1877^\circ,240^\circ),(79.1877^\circ,0^\circ),
\cr \ \ \cr \ \ &(37.3774^\circ,300^\circ),
(37.3774^\circ,180^\circ), (37.3774^\circ,60^\circ)].
\end{array}
\end{equation}
For future use in the calculation of (\ref{eq:W}), in the
construction of the matrix $D$ only the transducers located in the
five first positions were used.

\begin{figure}
 \includegraphics[width=12.5cm]{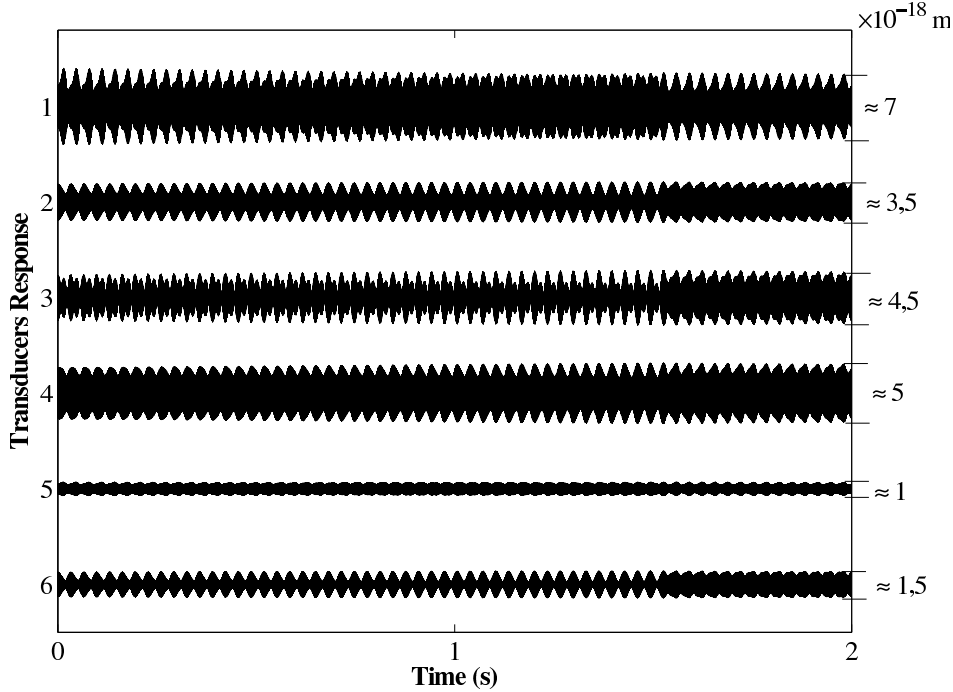}
  \caption{Transducers' outputs from the simulation using as input
   a g.w. emitted by a coalescing NS-NS system.}
  \label{fig:R}
\end{figure}

The values of $R^L$ were obtained using the software ``Schenberg
Simulator 2.0.0"\cite{cesartese}, initially in the absence of
noises. The mathematical model of the detector, which is the basis
for the software, can be found, for instance, in \cite{maga97b}. In
the software are used six two-mode transducers located on the
antenna's surface, distributed according to the angles listed above,
operating simultaneously. This software has proven effective in
solving the inverse problem\cite{cesartese}. The outputs of five of
these transducers (the same used to construct the matrix $D$) will
be used as the components of $R^L$.

If the equivalence between the spherical detectors and the
independent (cyclical) bars is correct, then the solution of the
inverse problem with the bars should yield the astrophysical
parameters of a source introduced as the input of the software that
simulates SCHENBERG's response. As will be shown, this equivalence
does exist.

The simulation that is going to be analyzed here was run with the
source represented by a waveform resultant from a coalescing neutron
star (NS) binary system\cite{duez01,duez01a}, incident on the
detector according to the angles $\Theta = 60^\circ$ and $\Phi =
45^\circ$ relative to the same system that defines the transducers'
positions. With the input from this source the software yielded the
outputs illustrated in Figure \ref{fig:R}.

\begin{figure}
 \includegraphics[width=8.5cm]{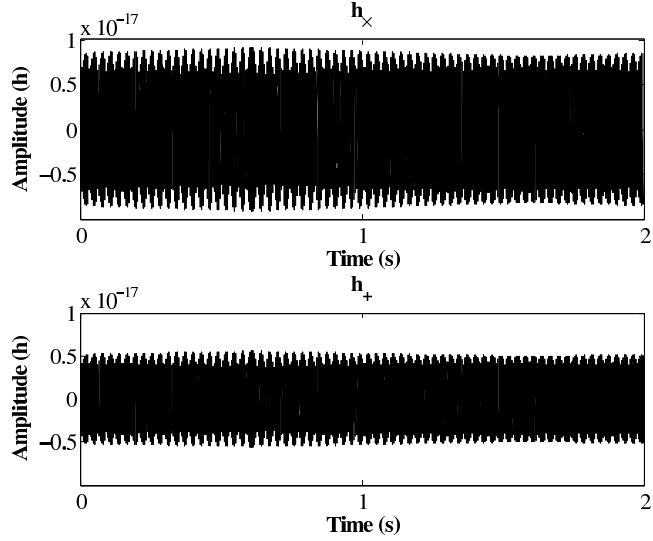}
  \caption{Amplitudes of the g.w. emitted by the simulated
  coalescing NS-NS system. These plots were obtained from the solution
  of the inverse problem with the method of independent bars
  using five outputs in the absence of noise.}
  \label{fig:ampl5}
\end{figure}

Using the first all the six outputs in eq. (\ref{eq:W}) the values
of $W$ were found and, consequently\cite{maga95}, the source's
direction was retrieved. In this particular case considered we have
a probability of $68.27\%$ that the direction is  $\Theta=60^\circ
\pm 2^\circ$ and $\Phi=44.4^\circ \pm1.4^\circ$. For the case of
five outputs in eq. (\ref{eq:W}) the result of the source's
direction is $\Theta=60^\circ \pm 2^\circ$ and $\Phi=44.2^\circ
\pm1.6^\circ$. It was verified that the result was in perfect
agreement with the source's input angles. The wave's amplitudes were
also determined, as shown in Figure \ref{fig:ampl5}.

Different combinations of five transducers were used, yielding the
same results as expected \cite{maga97b}. It became clear that the
method of independent bars using five outputs was as effective in
solving the inverse problem as the conventional method, which
considers all six transducers simultaneously.

\begin{figure}
 \includegraphics[width=10.5cm]{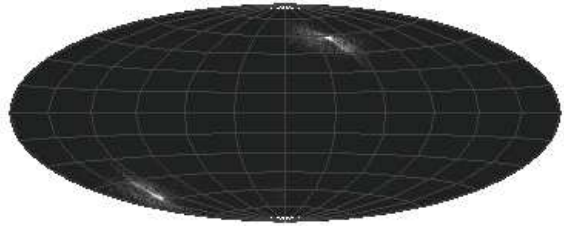}
  \vspace{-20mm}
  \caption{Source's direction as determined by the method of
  independent bars using five bars in the presence of Gaussian noise
  with SNR$\approx 18$. The different degrees of gray are related
  to the differences in the energy absorbed by the sphere.}
  \label{fig:hammer5noise}
\end{figure}

The procedure was repeated in the presence of Gaussian noises and
the results from both approaches coincided again. See Figure
\ref{fig:hammer5noise} for the case when the signal-to-noise ratio,
SNR, is approximately 18. In this figure two regions of the sky are
marked because there is a natural ambiguity in the determination of
the direction by only one detector; if more than one detector is
used then that there is a time delay between detections and one can
determine whether the wave came from up or down. The wider the
region marked, the higher the SNR.

The method of independent bars was also tested using the output of
only four transducers to solve the inverse problem in the presence
of Gaussian noise for the case when SNR is the approximately 93. The
preliminary results show that even using less transducers all the
four astrophysical parameters were determined from the method. The
result of the source's direction in this case is $\Theta=57^\circ
\pm 5^\circ$ and $\Phi=42^\circ \pm6^\circ$. This result take
obtained using the following transducers:
\begin{equation}
\begin{array}{cl}
[(\theta,\phi)] = &
[(79.1877^\circ,120^\circ),(79.1877^\circ,240^\circ),(79.1877^\circ,0^\circ),
\cr \ \ \cr \ \ & (37.3774^\circ,60^\circ)].
\end{array}
\end{equation}
However, in this case the computational time increases considerable.
Also, the precision of the results vary with the choice of the
transducers, confirming previous investigation \cite{maga97b}.

\section{Summary}

Presently, two spherical detectors exist (SCHENBERG and MiniGrail),
and they are expected to begin scientific runs soon. This work
presents results of a method that allows determination of
astrophysical parameters from the data of a spherical g.w. detector
using 6, 5 or 4 transducers. This is an original method, since in
the literature only 6 transducers were considered so far.

The research will continue with the solution of the inverse problem
using the method of independent bars with the output of 4 and 3
transducers, both assuming a noiseless and a noisy detector.
Circular and elliptical polarizations should be considered, in
addition to the linear ones. The detailed presentation of these
results will be published in a longer paper that is now in
preparation.

Also, it is intended to fully investigate SCHENBERG's directivity
pattern, including the simulation of real data from signals arriving
from different directions in the sky. This will allow statistical
estimates of the actual detector performance based on the
theoretical principles exposed here. Such an statistical analysis is
essential since the less transducers one has, the more important, if
not crucial, is to assess the significance of the detection.

\begin{acknowledgements}
The authors thank the financial support given by their respective
Brazilian funding agencies: CHL to CAPES, HABA to CNPq (grant \#
133228/2006-1), NSM and RMMJ to FAPESP (grants \# 2006/07316-0 and
\# 07/51783-4). A special acknowledgement is given to FAPESP for
supporting the construction and operation of the SCHENBERG detector
(grant \# 2006/56041-3, Thematic Project: ``New Physics in Space:
Gravitational Waves'').
\end{acknowledgements}

\newpage 

\clearpage

\end{document}